\documentclass[aip,pop,reprint,final,showpacs,groupedaddress,amsmath,amssymb]{revtex4-1}

\usepackage{graphicx}

\newcommand{\bew}{\begin{widetext}}
\newcommand{\eew}{\end{widetext}}
\newcommand{\beq}{\begin{equation}}
\newcommand{\eeq}{\end{equation}}
\newcommand{\bea}{\begin{eqnarray}}
\newcommand{\eea}{\end{eqnarray}}
\newcommand{\bes}{\begin{subequations}\begin{eqnarray}}
\newcommand{\ees}{\end{eqnarray}\end{subequations}}
\newcommand{\mbf}[1]{\mathbf{#1}}
\newcommand{\mrm}[1]{\mathrm{#1}}
\newcommand{\msf}[1]{\mathsf{#1}}

\newcommand{\epsi}{\epsilon}
\newcommand{\del}{\nabla}
\newcommand{\dotdel}{\cdot \nabla}
\newcommand{\divr}{\nabla \cdot}
\newcommand{\curl}{\nabla \times}
\newcommand{\dsub}[1]{\partial_{#1}}

\newcommand{\wt}[1]{\widetilde{#1}}
\newcommand{\abs}[1]{{\lvert {#1} \rvert}}

\newcommand{\uvec}[1]{{\hat{\mbf{#1}}}}
\newcommand{\dalem}{\square^2}
\newcommand{\expi}{\exp^i}
\newcommand{\real}{\mathrm{Re\,}}

\begin{document}

%%%%  RESET LINE NUMBER FONT
%%%%  AND MARGIN SPACING
\makeatletter
\@ifundefined{linenumbers}{}{
\renewcommand\linenumberfont{\normalfont\scriptsize}
\setlength\linenumbersep{0.5cm}
}
\makeatother

%\preprint{}

\title{Potential formulation of the dispersion relation for a uniform, magnetized plasma with stationary ions in terms of a vector phasor} %Title of paper



\author{Robert W. Johnson}
\email[]{robjohnson@alphawaveresearch.com}
\homepage[]{http://www.alphawaveresearch.com}
%\thanks{}
\affiliation{Alphawave Research, Jonesboro, GA 30238, USA}


%\date{\today}
\date{10 May 2012}

\begin{abstract}
The derivation of the helicon dispersion relation for a uniform plasma with stationary ions subject to a constant background magnetic field is reexamined in terms of the potential formulation of electrodynamics.  Under the same conditions considered by the standard derivation, the nonlinear self-coupling between the perturbed electron flow and the potential it generates is addressed.  The plane wave solution for general propagation vector is determined for all frequencies and expressed in terms of a vector phasor.  The behavior of the solution as described in vacuum units depends upon the ratio of conductivity to the magnitude of the background field.  Only at low conductivity and below the cyclotron frequency can significant propagation occur as determined by the ratio of skin depth to wavelength.
\end{abstract}


\maketitle %\maketitle must follow title, authors, abstract and \pacs


\section{Introduction}
In this article we reexamine the derivation of the helicon dispersion relation in terms of the potential formulation of electrodynamics.  Under the same approximations as used in the standard derivation, namely a uniform plasma with stationary ions subject to a constant background magnetic field with vanishing thermal stress and space charge density, the linearized equation of motion for the electron flow describes only the electron cyclotron resonance when the ion contribution to the material response is included through the friction term.  Only by addressing the nonlinear self-coupling of the perturbed electron flow to its own potential through the Lorentz term can a more interesting solution be found.  The plane wave solution for a general propagation vector is determined, whose frequency depends upon its inclination from the plane orthogonal to the direction of the background field.  These solutions are represented in terms of a vector of phase factors in addition to the oscillatory phase.  Several cases of interest are evaluated explicitly.  The main result is that propagation occurs only for frequencies in a range below the cyclotron frequency when the ratio of conductivity to the magnitude of the background magnetic field is sufficiently low that the skin depth exceeds the wavelength in the plasma.

The history of this derivation\cite{chen-137} goes back several decades.\cite{trivelpiece-1784,bowers-1961,legendy-1964,KMT-1965}  Chen surveys the experimental literature in his contribution\cite{chen-155r} to \textit{High Density Plasma Sources}, and with Boswell reviews the development of the theory throughout the twentieth century.\cite{chen-173,chen-174}  Recent examples have appeared in \textit{Physics of Plasmas} of the application of this theory to a toroidal vessel\cite{tripathi-697} and to an annularly bounded discharge chamber.\cite{yano-063501,yano-033510}  What these derivations have in common is their use of the electron equation of motion to determine the electric field rather than the electron flow as a consequence of their neglect of Gauss's law.  In the potential formulation of electrodynamics, the electric and magnetic fields are recognized as auxiliary expressions describing the spatial and temporal variations of the four-potential, which is determined by the inhomogeneous (source-bearing) Maxwell equations in conjunction with the gauge condition expressing the continuity of the potential.

This paper is organized as follows.  First we will review the standard derivation of the ``fast'' and ``slow'' helicon modes.  We will then reconsider the derivation in the potential formulation, including both the ion contribution to the material response and the nonlinear self-coupling of the electron flow to the potential it generates.  An example of the use of the potential formulation for the electrostatic case is given by Jankauskas and Kvedaras.\cite{jankauskas-274}  The solution of the material equation of motion is expressed in terms of a vector phasor describing the direction of the electron flow in addition to the scalar phasor describing its magnitude.  The solution of the potential equation of motion is expressed in terms of a complex propagation vector describing the wavelength and decay of oscillations at a given frequency.  These vectors are evaluated explicitly for a range of frequencies surrounding the electron cyclotron resonance, and the behavior of the solution is found to be determined by the ratio of the material's conductivity to the magnitude of the background magnetic field.  We will close by discussing how the theory must be extended before it can be applied to the description of an actual experimental configuration.

\section{Linearized Derivation}
First let us look at the derivation of the ``fast'' helicon mode, followed by the derivation of the coupled ``fast'' and ``slow'' modes.  The model for infinite conductivity is based on the field equations \bea
\curl \mbf{E}_\omega + \dsub{t} \mbf{B}_\omega = 0 \;,& \qquad \divr \mbf{B}_\omega = 0 \;, \label{eqn:homo} \\
\curl \mbf{B}_\omega = \mu_0 \mbf{J}_\omega \;,&  \label{eqn:ampere}
\eea and the linearized material equation of motion \beq \label{eqn:forE1}
\mbf{E}_\omega = \mbf{J}_\omega \times \mbf{B}_0 / e n_0 \;,
\eeq subject to the constraint $\divr \mbf{J}_\omega = 0$, where the subscript $\omega$ identifies the oscillating quantities, the constant background field $\mbf{B}_0 \equiv B_0 \uvec{z}$ defines a direction $\uvec{z}$, the uniform plasma density $n_0$ equals the number of electrons required for neutrality, and the plasma current $\mbf{J}_\omega = - e n_0 \mbf{V}_\omega$ for electron fluid velocity $\mbf{V}_\omega$.  The electric field is eliminated by the substitution \bes
- \dsub{t} \mbf{B}_\omega &=& \curl ( \mbf{J}_\omega \times \mbf{B}_0 ) / e n_0 \\
 &=& ( \mbf{B}_0 \dotdel ) \mbf{J}_\omega / e n_0 \;,
\ees where $\del n_0 \equiv 0$, and the current is eliminated by \beq
- \dsub{t} \mbf{B}_\omega = ( \mbf{B}_0 \dotdel ) ( \curl \mbf{B}_\omega ) / \mu_0 e n_0 \;,
\eeq which for a traveling wave with phase $\expi (\mbf{k} \cdot \mbf{r} - \omega t) \propto \expi (k_z z - \omega t)$, using the notation $\expi (\delta) \equiv e^{i \delta}$, yields the relation \beq \label{eqn:foralpha}
0 = ( \alpha - \curl ) \mbf{B}_\omega \;,
\eeq where the total wave number $\alpha = ( \omega / k_z ) ( \mu_0 e n_0 / B_0 )$ equals the magnitude of the propagation vector $\alpha = \abs{\mbf{k}}$, leading to the Helmholtz equation $(\del^2 + \alpha^2) \mbf{B}_\omega = 0$ describing the ``fast'' helicon mode.

For finite conductivity represented by a collision rate $\nu$ and including the effect of inertia represented by the electron mass $m_e$, the material equation of motion becomes \beq \label{eqn:forE2}
\mbf{E}_\omega = \mbf{J}_\omega ( \nu - i \omega ) m_e / e^2 n_0 + \mbf{J}_\omega \times \mbf{B}_0 / e n_0 \;,
\eeq whose curl leads to the relation \beq
0 = \left \lbrace \bigl [ (\omega + i \nu) / (k_z \omega_c) \bigr ] ( \curl )^2 - ( \curl ) + \alpha \right \rbrace \mbf{B}_\omega \;,
\eeq in terms of the electron cyclotron frequency $\omega_c \equiv e B_0 / m_e$, which can be factored\cite{KMT-1965} as \beq
0 = ( \beta_{+} - \curl ) ( \beta_{-} - \curl ) \mbf{B}_\omega \;.
\eeq  The two roots for nontrivial $\mbf{B}_\omega$, in terms of $\gamma \equiv k_z \omega_c / 2 (\omega + i \nu)$, give total wave numbers of \beq
\beta_\pm = \left [ 1 \pm ( 1 - 2 \alpha / \gamma )^{1/2} \right ] \gamma \;,
\eeq which are identified as the ``slow'' and ``fast'' modes, also known\cite{trivelpiece-1784,chen-175} as the Trivelpiece and Gould mode for $\beta_{+}$ and the helicon mode for $\beta_{-}$ respectively.

Let us now look at what Gauss's law has to say about these models, whose inclusion turns Eqns.~(\ref{eqn:homo}) and (\ref{eqn:ampere}) into the system for pre-Maxwell electrodynamics.  For either Eqn.~(\ref{eqn:forE1}) or Eqn.~(\ref{eqn:forE2}) one can write \bes
\divr \mbf{E}_\omega &=& \divr ( \mbf{J}_\omega \times \mbf{B}_0 ) / e n_0 \label{eqn:divE2} \\
 &=& \mbf{B}_0 \cdot ( \curl \mbf{J}_\omega ) / e n_0 \;,
\ees as the source generating the background field is external to the region of consideration, whereupon eliminating $\mbf{J}_\omega$ gives \bes
\divr \mbf{E}_\omega &=& \mbf{B}_0 \cdot [ (\curl)^2 \mbf{B}_\omega ] / \mu_0 e n_0 \\
 &=& - \del^2 ( \uvec{z} \cdot \mbf{B}_\omega ) ( B_0 / \mu_0 e n_0 ) \;.
\ees  For consistency with the approximation of neutrality ${\divr \mbf{E}_\omega} = 0$, one requires either $k^2 = 0$, where $k \equiv \abs{\mbf{k}}$, or $\uvec{z} \cdot \mbf{B}_\omega = 0$, which then implies $\mbf{J}_\omega = J_\omega \uvec{z}$ yielding $k_z = 0$.  For Eqn.~(\ref{eqn:forE1}), either condition results in the expression $\omega = 0$, equivalent to the statement\cite{griffiths-89} ``that the magnetic field is constant [] inside a perfect conductor.''  For Eqn.~(\ref{eqn:forE2}), Faraday's law can be written \bea
- \dsub{t} \mbf{B}_\omega &=& (\curl)^2 \mbf{B}_\omega ( \nu - i \omega ) m_e / \mu_0 e^2 n_0 \nonumber \\
 & & + \dsub{z} (  \curl \mbf{B}_\omega ) ( B_0 / \mu_0 e n_0 ) \;,
\eea which can be simplified to \beq
i \omega \mbf{B}_\omega = ( B_0 / \mu_0 e n_0 ) [ k^2 ( \nu - i \omega ) / \omega_c - k_z \mbf{k} \times ] \mbf{B}_\omega \;.
\eeq  Under the condition $k_z = 0$ one has the relation \bes
k_\perp^2 &=& i \mu_0 \omega \omega_c e n_0 / B_0 ( \nu - i \omega ) \\
 &=& i \mu_0 \omega \wt{\sigma} \;,
\ees in terms of the AC conductivity $\wt{\sigma} \equiv e^2 n_0 / m_e (\nu - i \omega)$, which one recognizes as the usual dispersion relation for a conductor\cite{griffiths-89} up to its neglect of the term $\mu_0 \epsi_0 \omega^2$ arising from the displacement current.  This result is consistent with the observation that the parallel current along $\uvec{z}$ is impervious to the effect of the linearized Lorentz term appearing in the material equation of motion.

The model as presented in the literature is quite specific about its neglect of the displacement current in the plasma region.  As Yano and Walker state,\cite{yano-033510} ``the displacement current in [the Maxwell-Ampere equation] is neglected for calculation of the plasma field, as it is always negligible in experiments.''  Let us consider the effect of its inclusion.  From Eqn.~(\ref{eqn:forE2}), the plasma current can be written as \bea
\mbf{J}_\omega &=& e n_0 \left [ ( \nu - i \omega ) m_e / e - \mbf{B}_0 \times \right ]^{-1} \cdot \mbf{E}_\omega \\
 &=& \wt{\sigma} \left [ \begin{array}{ccc} 1 & \xi & 0 \\ - \xi & 1 & 0 \\ 0 & 0 & 1 \end{array} \right ]^{-1} \cdot \mbf{E}_\omega \; ,
\eea where $\xi \equiv \omega_c / ( \nu - i \omega )$, which defines the gyrotropic conductivity tensor, \bea
\mbf{J}_\omega &=& \dfrac{\wt{\sigma}}{1 + \xi^2} \left [ \begin{array}{ccc} 1 & - \xi & 0 \\ \xi & 1 & 0 \\ 0 & 0 & 1 + \xi^2 \end{array} \right ] \cdot \mbf{E}_\omega \\
 &\equiv& \msf{\Sigma} \cdot \mbf{E}_\omega \; .
\eea  The Maxwell-Ampere equation can thus be written as \bea
\curl \mbf{B}_\omega &=& \mu_0 \epsi_0 \dsub{t} \mbf{E}_\omega + \mu_0 \mbf{J}_\omega \\
 &=& \mu_0 ( \epsi_0 \dsub{t} + \msf{\Sigma} ) \cdot \mbf{E}_\omega \; ,
\eea whose curl yields the relation \bea
- \del^2 \mbf{B}_\omega &=& \mu_0 ( \epsi_0 \dsub{t} + \msf{\Sigma} ) \cdot ( \curl \mbf{E}_\omega ) \\
 &=& - \mu_0 \dsub{t} ( \epsi_0 \dsub{t} + \msf{\Sigma} ) \cdot \mbf{B}_\omega \; ,
\eea whereupon substitution for harmonic oscillations gives \bea
0 &=& ( - k^2 + \omega^2 / c_0^2 + i \mu_0 \omega \msf{\Sigma} ) \cdot \mbf{B}_\omega \\
 &\equiv& \msf{\Upsilon} \cdot \mbf{B}_\omega \; ,
\eea for $\mu_0 \epsi_0 c_0^2 \equiv 1$.  For nontrivial $\mbf{B}_\omega$, one requires the matrix $\Upsilon$ to be singular (non-invertible), thus its determinant must vanish $\det \Upsilon = 0$, yielding the dispersion relation \bea
0 &=& \left ( i \mu_0 \omega \wt{\sigma} + \dfrac{\omega^2}{c_0^2} - k^2 \right ) \nonumber \\
 & & \left [ \left ( \dfrac{i \mu_0 \omega \wt{\sigma}}{1+\xi^2} + \dfrac{\omega^2}{c_0^2} - k^2 \right )^2 - \left ( \dfrac{\mu_0 \omega \wt{\sigma} \xi}{1+\xi^2} \right )^2 \right ] \; ,
\eea which has three positive solutions indexed by $\eta \in [-1, 0, 1]$ and expressed as \beq
k^2_\eta = \omega^2 / c_0^2 + i \mu_0 \omega \wt{\sigma} / (1 + i \xi \eta) \; .
\eeq  The solution $k_0$ is commonly identified\cite{dendybook-93} as the ordinary mode, and the solutions $k_\pm$ as the extraordinary modes.

The approximation of neutrality, according to Eqn.~(\ref{eqn:divE2}), now requires \beq
0 \propto \left ( -\del^2 + \dsub{t}^2 / c_0^2 \right ) ( \uvec{z} \cdot \mbf{B}_\omega ) \; ,
\eeq where the displacement term has contributed to the first factor.  The condition $\mbf{B}_\omega \perp \uvec{z}$ requires $(\epsi_0 \dsub{t} + \msf{\Sigma}) \cdot \mbf{E}_\omega \parallel \uvec{z}$, thus $\mbf{E}_\omega \parallel \uvec{z}$ and $\mbf{k} \perp \uvec{z}$ as before.  The condition $k_\eta^2 = \omega^2 / c_0^2$ can be satisfied by $\wt{\sigma} \rightarrow 0$, \textit{i.e.} the vacuum, or for the extraordinary modes $k_\pm$ at the frequency $\omega = - i \nu$, which describes not an oscillation but a solution that decays exponentially with time, $e^{- i \omega t} \rightarrow e^{- \nu t}$.  Note that the satisfaction of Gauss's law for a neutral medium is what prevents these models from supporting propagation with a component parallel to the background field.

\section{Nonlinear Derivation}
Our first objection to the preceding derivations of the helicon dispersion relation is mathematical, as the linearized form of the Lorentz term neglects the self-interaction between the electron fluid and the potential it generates.  In order to describe properly the phenomenon of resonance within the electron fluid, one needs to account for the influence one fluid element has on another.  That effect appears formally as a nonlinear contribution to the Lorentz force which can dominate the dynamics when the background field is weak compared to that generated by source currents within the plasma.

Our next objection is not so much one of mathematics but rather one of physics, in that the equation chosen to determine the electric field, either Eqn.~(\ref{eqn:forE1}) or Eqn.~(\ref{eqn:forE2}), does not tell the whole story of the material response.  The specification of charge neutrality requires the presence of a positively charged ion background in addition to the negatively charged electron fluid.  Just because its fluid velocity vanishes does not mean that its equation of motion is worthless, as the ion background must still satisfy the equation for its contribution to momentum conservation.

Our last objection to the prevailing derivations is that the abbreviated set of Maxwell equations found in Eqns.~(\ref{eqn:homo}-\ref{eqn:ampere}) does not satisfy the formalism of the potential formulation of electrodynamic field theory.\cite{ryder-qft,naka-798212}  Conspicuous by its absence is Gauss's law, which describes the fundamental relation between a source and the field it generates and is equivalent to the Maxwell-Ampere equation under a Lorentz transformation.  Also absent is the term for displacement current vital to the description of electromagnetic oscillations as well as the covariant expression of the continuity of the source terms.\cite{bork-854,griffiths-89}  Let us now reconsider the derivation of the dispersion relation in the potential formulation, based upon the complete classical Maxwell system, for a model which includes both the contribution from the ion equation of motion as well as the nonlinear self-interaction of the electron fluid.

\subsection{Model equations}
For each species $s \in \{i,e\}$, continuity is expressed through the total derivative of density defined by \beq
\dot{n}_s \equiv \dsub{t} n_s + \divr ( n_s \mbf{V}_s ) \;,
\eeq where $\mbf{V}_s$ is the species flow velocity and $\dot{n}_s$ is the particle source rate, and the convective derivative of velocity is \beq
\dot{\mbf{V}}_s \equiv \dsub{t} \mbf{V}_s + ( \mbf{V}_s \dotdel ) \mbf{V}_s \;.
\eeq  Considering a singly ionized species such that $e_i = e$ and $e_e = -e$ under the assumption of vanishing ion flow $\mbf{V}_i \equiv 0$, both the net current and net momentum are proportional to the electron flux, $n_e \mbf{V}_e = - \mbf{J} / e = \mbf{K} / m_e$, thus the fluid acceleration can be written $\dot{\mbf{K}} = m_e (\dot{n}_e \mbf{V}_e + n_e \dot{\mbf{V}}_e)$.  The statements $\dsub{t} \mbf{V}_i = 0$ and $\dsub{t} n_i = 0$ imply that the ions provide a fixed background for the electron dynamics.  The equations of motion\cite{dendybook-93,staceybook05} in the absence of particle sources $\dot{n}_s = 0$ are thus \bea
\divr ( n_i \msf{T}_i ) &=& \mbf{F}_{ie} + e n_i \mbf{E} \; , \label{eqn:ion} \\
m_e n_e \dot{\mbf{V}}_e + \divr ( n_e \msf{T}_e ) &=& \mbf{F}_{ei} - e n_e ( \mbf{E} + \mbf{V}_e \times \mbf{B} ) \;, \label{eqn:eon}
\eea where $n_s \msf{T}_s$ is the thermal stress tensor for species $s$.  Just because the ion flow vanishes does not mean that its equation is worthless, as the electron flow makes an appearance through the friction term, $\mbf{F}_{ie} \equiv - \mbf{F}_{ei} = m_e n_e \nu_{ei} \mbf{V}_e$ for interspecies collision rate $\nu_{ei}$.  Summing Eqns.~(\ref{eqn:ion}) and (\ref{eqn:eon}) yields \beq
m_e n_e \dot{\mbf{V}}_e + \divr ( n_i \msf{T}_i + n_e \msf{T}_e ) = e ( n_i - n_e ) \mbf{E} - e n_e \mbf{V}_e \times \mbf{B} \;, \label{eqn:forVe}
\eeq which is simply the statement of net momentum conservation in the plasma under the given conditions, \beq \label{eqn:netforce}
\dot{\mbf{K}} + \divr ( n \msf{T} ) = j \mbf{E} + \mbf{J} \times \mbf{B} \;,
\eeq where $j = \sum_s n_s e_s$ is the net charge density and $n T = \sum_s n_s \msf{T}_s$ is the net thermal stress.  Under the approximation of uniform thermal stress $\divr (n_s \msf{T}_s) = 0$ and enforcing neutrality $n_e = n_i$ such that $\divr (n_e \mbf{V}_e) = 0$, one is left with the system of equations \bea
- m_e \nu_{ei} \mbf{V}_e &=& e \mbf{E} \;, \label{eqn:ioneqn} \\
m_e ( \dsub{t} + \mbf{V}_e \dotdel ) \mbf{V}_e &=& e \mbf{B} \times \mbf{V}_e \;, \label{eqn:eoneqn}
\eea which displays the electron fluid's coupling to inertia through $m_e$ on the LHS and to electromagnetism through $e$ on the RHS.  Only by satisfying both equations of motion does one achieve a consistent theory for the material response.

Turning now to the electromagnetic sector, the homogeneous Maxwell equations are satisfied identically in the potential formulation $\mbf{B} \equiv \curl \mbf{A}$ and $\mbf{E} \equiv - \del \Phi - \dsub{t} \mbf{A}$, thus they cannot determine any physically relevant degrees of freedom.  The physical relation between the potential and its source is given by the inhomogeneous Maxwell equations \beq \label{eqn:inhomo}
\curl \mbf{B} - \mu_0 \epsi_0 \dsub{t} \mbf{E}  = \mu_0 \mbf{J} \;, \qquad \divr \mbf{E} = j / \epsi_0 \;,
\eeq where the inclusion of the displacement current is essential to the description of electromagnetic propagation as well as the covariant expression of charge conservation.  Selecting the Lorenz gauge $\dsub{t} a + \divr \mbf{A} = 0$, where the scalar potential $a \equiv \Phi / c_0^2$, the continuity of the potential mirrors that of the source $\dsub{t} j + \divr \mbf{J} = 0$, and the field equations are covariant $- \dalem A^\nu = \mu_0 J^\nu$, where the d'Alembertian operator is defined as $\dalem \equiv \del^2 - c_0^{-2} \dsub{t}^2$.  For a neutral medium $j \equiv 0$ the scalar potential vanishes $a \equiv 0$ (technically is constant), thus one is left with the gauge condition $\divr \mbf{A} = 0$ and the field equation \beq
- \del^2 \mbf{A} + c_0^{-2} \dsub{t}^2 \mbf{A}  = \mu_0 \mbf{J} \;,
\eeq expressing the relation between the vector potential $\mbf{A}$ and its source current $\mbf{J}$.

\subsection{Linearized solution}
Considering a region with uniform plasma density $\del n_e \equiv 0$ such that $\divr \mbf{V}_e = 0$ and subject to a constant background magnetic field $\mbf{B}_0$ along $\uvec{z}$, the linearized equation of motion reads \beq \label{eqn:linform}
\dsub{t} \mbf{V}_e = (e/m_e) \mbf{B}_0 \times \mbf{V}_e \; .
\eeq  Nothing defined in the model breaks translational invariance, so let us work in Cartesian coordinates with position vector $\mbf{r} = (x, y, z)$, and let us write the phasor for the electron flow $\mbf{V}_e =  \real \wt{\mbf{V}}_e$ as $\wt{\mbf{V}}_e = \expi (- \omega t) (\wt{V}_x, \wt{V}_y, \wt{V}_z)$.  The phasor equation of motion in terms of the cyclotron frequency $\omega_c$ is then \beq
( \wt{V}_x, \wt{V}_y, \wt{V}_z ) = i ( \omega_c / \omega ) ( -\wt{V}_y, \wt{V}_x, 0 ) \;,
\eeq thus the flow along $\uvec{z}$ must vanish $\wt{V}_z = 0$, and the relations $\wt{V}_x / \wt{V}_y = - i ( \omega_c / \omega )$ and $\wt{V}_y / \wt{V}_x = i ( \omega_c / \omega )$ must hold independently.  For non-vanishing flow, one can then write \bes
0 &=& 1 - (\wt{V}_x / \wt{V}_y) (\wt{V}_y / \wt{V}_x) \\
 &=& 1 - ( \omega_c / \omega )^2 \;,
\ees with the positive root $\omega = \omega_c$ which describes the material response at the electron cyclotron resonance.  Having assumed away most of the relevant physics, there is nothing else left for that equation to describe.  The phasor solution can thus be written in terms of a scalar phasor $\expi (\delta)$ for the magnitude and a vector phasor $\wt{\mbf{f}}$ for the direction, \bes
\wt{\mbf{V}}_e &\equiv& V \expi (\delta) \wt{\mbf{f}} \\
 &=& V \expi (- \omega_c t) (1, i, 0) \;,
\ees where $V$ is a real constant bearing units of velocity and $\wt{\mbf{f}} \equiv \mbf{f}_{+} + i \mbf{f}_{-}$ is the complex sum of real unit vectors satisfying the requirements of normalization $\abs{\mbf{f}_{+}} = \abs{\mbf{f}_{-}} = 1$ and orthogonality $\mbf{f}_{+} \cdot \mbf{f}_{-} = 0$.  Note that this solution does not describe a propagating wave as the entire region is oscillating at the same phase in time, having no dependence on position $\mbf{r}$.  The vector phasor $\wt{\mbf{f}}$ describes the direction of the flow at two times separated by one quarter of a cycle with a relative phase $\expi (-\pi / 2) = - i$.

\subsection{Nonlinear solution}
Let us now consider the coupled nonlinear system of equations \bea
\dsub{t} \wt{\mbf{V}}_e + ( \wt{\mbf{V}}_e \dotdel ) \wt{\mbf{V}}_e &=& (e/m_e) (\mbf{B}_0 + \curl \wt{\mbf{A}}) \times \wt{\mbf{V}}_e \; , \label{eqn:forV} \\
- \del^2 \wt{\mbf{A}} + c_0^{-2} \dsub{t}^2 \wt{\mbf{A}} &=& - \mu_0 e n_e \wt{\mbf{V}}_e \; , \label{eqn:forA}
\eea where the phasor for the generated potential $\wt{\mbf{A}} \equiv (A/V) \expi (\delta_A) \wt{\mbf{V}}_e$ has a phase of $\delta_A$ relative to $\wt{\mbf{V}}_e$, and seek solutions for $\wt{\mbf{V}}_e$ and $\wt{\mbf{A}}$ under the constraints \bea
\divr \wt{\mbf{V}}_e = 0 \; , \quad \divr \wt{\mbf{A}} &=& 0 \; , \label{eqn:divcon} \\
\dsub{t} \wt{\mbf{A}} - (m_e \nu_{ei} / e) \wt{\mbf{V}}_e &=& 0 \; , \label{eqn:Econ}
\eea where the final constraint is Eqn.~(\ref{eqn:ioneqn}) in terms of the potential.  The oscillatory phase $\delta = \wt{\mbf{k}} \cdot \mbf{r} - \omega t$ now allows for variation in space through the complex propagation vector $\wt{\mbf{k}} \equiv \mbf{k}_{+} + i \mbf{k}_{-}$ whose real part $\mbf{k}_{+}$ determines the oscillation's wavelength and whose imaginary part $\mbf{k}_{-}$ describes its attenuation.  The propagation vector can also be written in terms of its direction, \beq
\uvec{k} \equiv (\cos \theta_k \cos \phi_k ,\sin \theta_k  \cos \phi_k ,\sin \phi_k) \;,
\eeq and magnitude $\wt{k} \equiv k \expi (\delta_k)$ as $\wt{\mbf{k}} \equiv \wt{k} \uvec{k}$, where the angle $\theta_k \in [-\pi, \pi)$ is its azimuth from $\uvec{x}$, and the angle $\phi_k \in [-\pi/2, \pi/2]$ is its inclination from the plane whose normal is $\uvec{z}$.  In these coordinates $\del \rightarrow i \wt{\mbf{k}}$, and the divergence constraints in Eqns.~(\ref{eqn:divcon}) imply $\wt{\mbf{k}} \cdot \wt{\mbf{f}} = 0$ so that the convective acceleration vanishes $( \wt{\mbf{V}}_e \dotdel ) \wt{\mbf{V}}_e = 0$.

\begin{figure}
\includegraphics[]{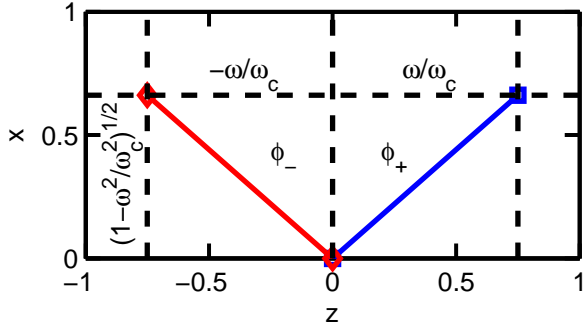}%
\caption{\label{fig:A}(Color online.)  The propagation vector $\wt{\mbf{k}}$ has an inclination from the plane orthogonal to $\uvec{z}$ of $\phi_\pm$ determined by the ratio of the oscillating frequency $\omega$ to the electron cyclotron frequency $\omega_c$.  Shown are the vectors $\uvec{k}$ for $\phi_{+}$ as $\square$ and $\phi_{-}$ as $\lozenge$ when $\omega / \omega_c = 3/4$.}
\end{figure}

The divergence constraints can be satisfied constructively by specifying $\wt{f}_z = - ( \wt{k}_x \wt{f}_x + \wt{k}_y \wt{f}_y ) / \wt{k}_z$ for both $\wt{\mbf{V}}_e$ and $\wt{\mbf{A}}$.  The remaining constraint requires $\omega \wt{\mbf{A}} = i ( m_e \nu_{ei} / e ) \wt{\mbf{V}}_e$, thus determining $\delta_A = \pi / 2$ and $A/V = m_e \nu_{ei} / e \omega$.  In this formulation, the ion contribution Eqn.~(\ref{eqn:Econ}) plays the role of Ohm's law, and the electron contribution Eqn.~(\ref{eqn:forV}), rewritten with the nonlinear term isolated as \beq \label{eqn:forAxV}
\left[ (m_e/e) \dsub{t} - \mbf{B}_0 \times \right] \wt{\mbf{V}}_e  = (\curl \wt{\mbf{A}}) \times \wt{\mbf{V}}_e \; ,
\eeq describes the oscillatory equilibrium in terms of momentum balance.  One factor of $e^{i \delta}$ appears on either side of that equation, but a second one appears on the RHS, which can be rewritten as $(\curl \wt{\mbf{A}}) \times \wt{\mbf{V}}_e = \wt{\gamma} \expi(2 \delta) \wt{\mbf{k}}$, where $\wt{\gamma} = A V [ \begin{array}{cc} \wt{f}_x & \wt{f}_y \end{array} ] \msf{\Gamma} [ \begin{array}{cc} \wt{f}_x & \wt{f}_y \end{array} ]^T$ in terms of \beq
\msf{\Gamma} = \left[ \begin{array}{cc} 1 + \cos^2 \theta_k \cot^2 \phi_k & \cos \theta_k \sin \theta_k \cot^2 \phi_k \\
\cos \theta_k \sin \theta_k \cot^2 \phi_k & 1 + \sin^2 \theta_k \cot^2 \phi_k \end{array} \right] \; .
\eeq  The common factor of $\wt{\gamma}$ lets one use the $\uvec{z}$ component of Eqn.~(\ref{eqn:forAxV}), \beq
(m_e/e) \dsub{t} \wt{V}_z = \wt{\gamma} \expi(2 \delta) \wt{k}_z \; ,
\eeq to simplify the $\uvec{x}$ and $\uvec{y}$ components into a system of equations for $\wt{f}_x$ and $\wt{f}_y$, \bea
\wt{f}_x / \wt{f}_y &=& - \dfrac{ \cos \theta_k \sin \theta_k \cot^2 \phi_k + i  (\omega_c/\omega) }{ 1 + \cos^2 \theta_k \cot^2 \phi_k } \; , \\
\wt{f}_y / \wt{f}_x &=& - \dfrac{ \cos \theta_k \sin \theta_k \cot^2 \phi_k - i  (\omega_c/\omega) }{ 1 + \sin^2 \theta_k \cot^2 \phi_k } \; ,
\eea which one can solve for $\phi_k (\omega)$ by writing \bes
0 &=& 1 - (\wt{f}_x / \wt{f}_y) (\wt{f}_y / \wt{f}_x) \\
 &\propto& (\omega_c \sin \phi_k - \omega) (\omega_c \sin \phi_k + \omega) \;,
\ees with the solutions $\phi_\pm = \pm \arcsin (\omega / \omega_c)$ as shown in Fig.~\ref{fig:A}, which says that propagation at a frequency $\omega \lesssim \omega_c$ is mostly along $ \pm \uvec{z}$ while at low frequency $0 \lesssim \omega$ is mostly perpendicular to $\uvec{z}$.  The ratio \beq
\wt{f}_z / \wt{f}_x = - \dfrac{ [ \cos \theta_k + i (\omega_c/\omega) \sin \theta_k ] \cot \phi_k }{ 1 + \sin^2 \theta_k \cot^2 \phi_k }
\eeq in terms of $\phi_\pm$ such that $\cot \phi_\pm = \pm (\omega_c^2 / \omega^2 - 1)^{1/2}$  lets one express the vector phasor as \beq \label{eqn:forf}
\wt{\mbf{f}} \propto \left[ \begin{array}{c} 1 + ( \omega_c^2 / \omega^2 - 1 ) \sin^2 \theta_k \\ ( 1 - \omega_c^2 / \omega^2 ) \cos \theta_k \sin \theta_k + i ( \omega_c / \omega ) \\ \mp ( \omega_c^2 / \omega^2 - 1 )^{1/2} [ \cos \theta_k + i (\omega_c/\omega)\sin \theta_k ] \end{array} \right] \; ,
\eeq where the constant of proportionality normalizes both the real and imaginary components of $\wt{\mbf{f}}$, which is a primary result of this investigation.  One can verify that $\hat{\mbf{k}} \cdot \wt{\mbf{f}} = 0$ so that continuity is preserved.

Let us examine two specific cases of interest.  Suppose first that $\delta_1 = \wt{k} z - \omega t$.  The explicit solution of Eqn.~(\ref{eqn:forAxV}) recovers the electron cyclotron resonance $\wt{\mbf{f}}_1 = (1, i, 0)$ for $\omega_1 = \omega_c$ such that $\phi_1 = \pi / 2$.  Now suppose that $\delta_2 = \wt{k} ( x + y + z ) /  3^{1/2} - \omega t$, where the square root of 3 normalizes the unit vector.  The explicit solution now gives $\wt{\mbf{f}}_2 \propto (1, e^{i 2 \pi / 3}, e^{-i 2 \pi / 3})$ for $\omega_2 = \omega_c / 3^{1/2}$ such that $\phi_2 = \arcsin 3^{-1/2}$ and $\theta_2 = \pi / 4$.  One can check that Eqn.~(\ref{eqn:forf}) reproduces these phase factors using the appropriate angles for the propagation vector.

\begin{figure}
\includegraphics[scale=0.8]{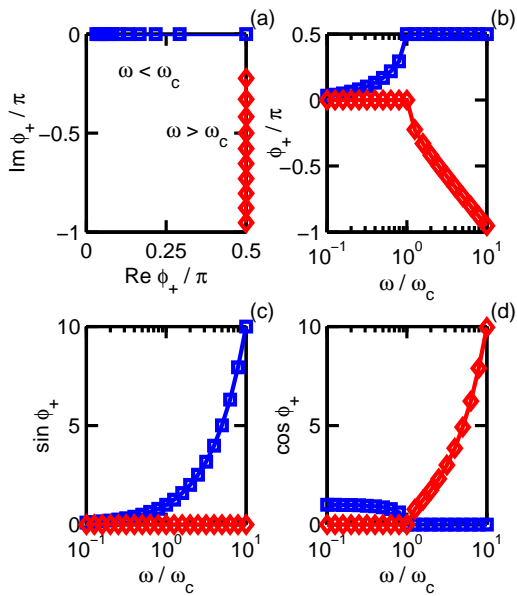}%
\caption{\label{fig:B}(Color online.)  Shown are the real $\square$  and imaginary $\lozenge$ components of $\phi_{+}$ as a function of the frequency ratio $\omega / \omega_c$ in (a) and (b), as well as its sine in (c) and its cosine in (d).}
\end{figure}

The trigonometric functions extend analytically to the complex plane $\phi \rightarrow \wt{\phi}$ such that $\cos^2 \wt{\phi} + \sin^2 \wt{\phi} = 1$ is preserved.\cite{flanigan-1983}  In particular, one can extend the expression $\sin \phi_\pm = \pm \omega / \omega_c$ to the domain $\omega > \omega_c$ such that $\cos \phi_\pm = i (\omega^2 / \omega_c^2 - 1)^{1/2}$.  The evolution of the complex $\phi_{+}$ as a function of the frequency ratio $\omega / \omega_c$ is depicted in Fig.~\ref{fig:B}, as are its sine and cosine.  The angle $\phi_{+}$ proceeds along the real axis from 0 to $\pi/2$ as $\omega$ goes from 0 to $\omega_c$, and then it acquires an imaginary component when $\omega > \omega_c$.  The sine is entirely real, while the cosine goes from being real to imaginary as the frequency ratio exceeds unity; consequently, the unit vector $\uvec{k}$ is itself complex in this regime.  Nonetheless, it maintains its normalization $\uvec{k} \cdot \uvec{k} = 1$ and its orthogonality to the vector phasor $\uvec{k} \cdot \wt{\mbf{f}} = 0$.  We will come back to the interpretation of $\wt{\mbf{k}}$ after we solve the conductive d'Alembertian equation for its magnitude $\wt{k}$.

What remains is to solve Eqn.~(\ref{eqn:forA}) for $\wt{k} (\omega)$, which is a standard problem in electro\-dynamics.\cite{griffiths-89}  Having written the propagation vector in terms of its magnitude and direction, the d'Alembertian operator simplifies considerably, $- \dalem = \wt{k}^2 - \omega^2 / c_0^2$.  Rearranging factors after writing $\wt{\mbf{V}}_e$ in terms of $\wt{\mbf{A}}$ yields the scalar relation $- \dalem = i \mu_0 \sigma \omega$ for DC conductivity $\sigma \equiv e^2 n_e / m_e \nu_{ei}$, which has the positive root \beq \label{eqn:fork}
\wt{k} = (\omega / c_0) (1 + i \sigma / \epsi_0 \omega)^{1/2}
\eeq with magnitude $k = (\omega / c_0) (1 + \sigma^2 / \epsi_0^2 \omega^2)^{1/4}$ and phase $\delta_k = \arctan (\sigma / \epsi_0 \omega) / 2$.  When $\omega \leq \omega_c$ the real propagation and decay vectors $\mbf{k}_{+}$ and $\mbf{k}_{-}$ point in the same direction such that $\wt{k} = \abs{\mbf{k}_{+}} + i \abs{\mbf{k}_{-}}$, but when $\omega > \omega_c$ the situation gets more complicated because of the complex form of $\uvec{k} \equiv \uvec{k}_{+} + i \uvec{k}_{-}$, where $\abs{\uvec{k}_{+}}^2 = \omega^2 / \omega_c^2$ and  $\abs{\uvec{k}_{-}}^2 = \omega^2 / \omega_c^2 - 1$.  In this notation, $\uvec{k}_{+} \neq \mbf{k}_{+} / \abs{\mbf{k}_{+}}$ as each expression refers to a different object; $\uvec{k}_{+}$ is the real part of $\uvec{k}$, whereas $\mbf{k}_{+} / \abs{\mbf{k}_{+}}$ gives the direction of propagation according to the real part of $\wt{\mbf{k}}$.  The direction of decay is given by $\mbf{k}_{-} / \abs{\mbf{k}_{-}}$, which is not aligned with the direction of propagation when the frequency ratio exceeds unity.  Recalling $\wt{\mbf{k}} \equiv \mbf{k}_{+} + i \mbf{k}_{-}$ such that \beq \label{eqn:fordecay}
\expi (\wt{\mbf{k}} \cdot \mbf{r}) = \exp (-\mbf{k}_{-} \cdot \mbf{r}) \expi (\mbf{k}_{+} \cdot \mbf{r}) \;,
\eeq the complex propagation vector describes the dispersion of electromagnetic radiation in a conductive medium with skin depth $\lambda_\sigma = 1 / \abs{\mbf{k}_{-}}$, wavelength $\lambda_\omega = 2 \pi / \abs{\mbf{k}_{+}}$, phase speed $c_\omega = \omega / \abs{\mbf{k}_{+}}$, and group speed $v_\omega = \partial \omega / \partial \abs{\mbf{k}_{+}}$.

\section{Evaluation of the Nonlinear Solution}
The evaluation of the solution to the system of equations as a function of the frequency ratio $\omega / \omega_c$ divides naturally into that for the vector phasor described by $\wt{\mbf{f}}$ and for the scalar phasor described by $\wt{\mbf{k}}$.  The solution is translationally invariant, as it is expressed in terms of its magnitude $V$ at some arbitrary location identified as the origin of the coordinate system.  The expression for $\wt{\mbf{f}}$ in Eqn.~(\ref{eqn:forf}) depends only upon the frequency ratio and the arbitrary azimuth $\theta_k$, whereas the expression for $\wt{\mbf{k}}$ depends additionally upon the conductivity $\sigma$.  Let us now look at each in turn.

\subsection{Evaluation of the vector phasor}
The vector phasor $\wt{\mbf{f}} \equiv \mbf{f}_{+} + i \mbf{f}_{-}$ describes the direction of the electron flow at two times separated by one quarter of a cycle.  Under the conditions of normalization and orthogonality it can be reduced to two scalar degrees of freedom corresponding to two of the components of Eqn.~(\ref{eqn:forAxV}), but it is instructive to evaluate all of its components explicitly.  The third component of Eqn.~(\ref{eqn:forAxV}) determines the direction $\uvec{k}$ through its dependence on $\phi_\pm$ as a function of the frequency ratio $\omega / \omega_c$.

\begin{figure}
\includegraphics[scale=0.8]{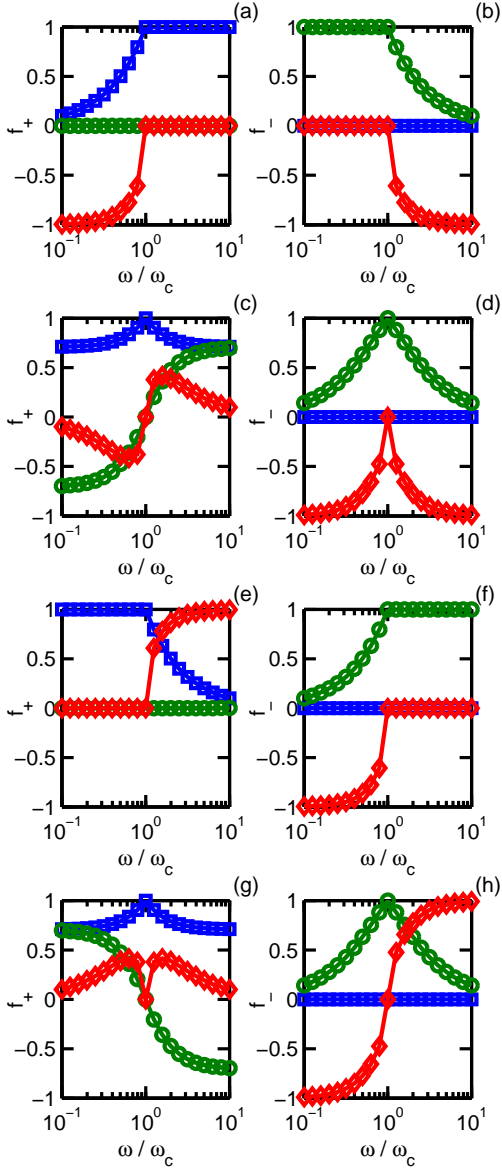}%
\caption{\label{fig:C}(Color online.)  The vector phasor $\wt{\mbf{f}}$, here for $\phi_{+}$, depends upon the frequency ratio $\omega / \omega_c$ and the propagation azimuth $\theta_k$, equal to 0 in (a) and (b), $\pi/4$ in (c) and (d), $\pi/2$ in (e) and (f), and $3\pi/4$ in (g) and (h).  On the left is its real part $\mbf{f}_{+}$, and on the right is its imaginary part $\mbf{f}_{-}$.  The $(x,y,z)$ components are displayed as $\square$, $\bigcirc$, and $\lozenge$ respectively.}
\end{figure}

Shown in Fig.~\ref{fig:C} are the $(x,y,z)$ components of $\mbf{f}_{+}$ and $\mbf{f}_{-}$ as a function of $\omega / \omega_c$ for various azimuths $\theta_k$ spanning one half revolution and selecting $\phi_{+}$.  The components of $\uvec{k}$ can be inferred from Fig.~\ref{fig:B}.  Focusing on the case of $\theta_k = 0$ such that $\uvec{k}$ has no $\uvec{y}$ component, the flow is primarily in the $y \mbox{-} z$ plane for $\omega \ll \omega_c$, in the $x \mbox{-} y$ plane for $\omega = \omega_c$, and in the $x \mbox{-} z$ plane for $\omega \gg \omega_c$.  Despite having flow in the same plane as propagation when the frequency ratio exceeds unity, one can verify that the divergence constraint $\uvec{k} \cdot \wt{\mbf{f}} = 0$ is satisfied.

\subsection{Evaluation of the scalar phasor}
The scalar phasor $\expi (\delta)$ for $\delta = \wt{\mbf{k}} \cdot \mbf{r} - \omega t$ describes the behavior of the time harmonic solution in space through the complex propagation vector $\wt{\mbf{k}}$ as a function of the frequency ratio $\omega / \omega_c$ and propagation azimuth $\theta_k$, as well as the plasma conductivity $\sigma$.  In the following let us set $\theta_k = 0$ so that propagation and decay are solely in the $x \mbox{-} z$ plane, and let us select the $\phi_{+}$ solution.  The electromagnetic sector reduces to the scalar equation $- \dalem = i \mu_0 \sigma \omega$ which determines $\wt{k}$ according to Eqn.~(\ref{eqn:fork}), thus we need to consider what is a reasonable range for the parameter $\sigma$ in units of siemens per meter.

\begin{figure}
\includegraphics[scale=0.8]{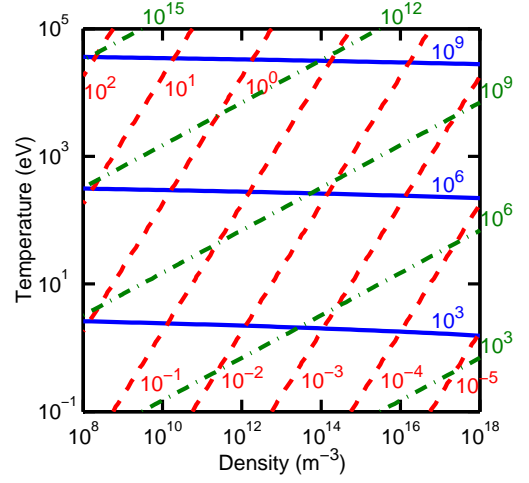}%
\caption{\label{fig:D}(Color online.)  Shown are contours for conductivity $\sigma$ (solid), Debye length $\lambda_D$ (dashed), and Debye density $n_D$ (dash-dot) over the $n \mbox{-} T$ plane with temperature expressed in units of electron-volts.  The contours for $\sigma$ are labeled vertically, for $\lambda_D$ are horizontal, and for $n_D$ are to the outside of the plot.}
\end{figure}

The conductivity is proportional to the ratio of density to collision frequency $\sigma \propto n_e / \nu_{ei}$, where $\nu_{ei}$ depends upon both density and temperature.  The range of density and temperature spanned by matter in the plasma state covers many orders of magnitude for each.  A common parametrization of the $n \mbox{-} T$ plane, where $T$ is the thermal energy, is given by the (electron) Debye length $\lambda_D \equiv (\epsi_0 T_e / n_e e^2 )^{1/2}$ and Debye density $n_D \equiv (4 \pi / 3) \lambda_D^3 n_e$.  Their contours are shown in Fig.~\ref{fig:D} over a reasonable range of the $n \mbox{-} T$ plane.  Also shown are contours for the conductivity evaluated as follows.  For a gaseous plasma\cite{staceybook05} one can estimate the collision rate from the expression \beq
\nu_{ei}^{-1} = 6 \epsi_0^2 ( 6 \pi m_e T_e^3 )^{1/2} / ( n_e e^4 \log \Lambda ) \; ,
\eeq where $\Lambda = 12 \pi [(\epsi_0 T_e / e^2)^3 / n_e]^{1/2}$.  One can see that the conductivity $\sigma$ has only a weak dependence on density $n_e$ through $\log \Lambda$.  A reasonable range for $\sigma$ can thus be estimated as $10^0 < \sigma < 10^{10}$ which spans materials from poorly to highly conductive.

The expression for $\wt{k}$ depends explicitly on the ratio of conduction to displacement current $\rho \equiv \sigma / \epsi_0 \omega = \tan (2 \delta_k)$.  In the presence of a background magnetic field $\mbf{B}_0$ with magnitude $B_0$ in units of tesla, that ratio can be rewritten in terms of the cyclotron frequency as $\rho = (\omega_c / \omega) \rho_0$, where \beq
\rho_0  \equiv \sigma / \epsi_0 \omega_c
\eeq is a unitless parameter depending on the conductivity and the magnitude of the background field.  The value of $\rho_0$ is ultimately what determines the dispersion of electromagnetic radiation in a conductive medium subject to an external magnetic field.

To determine the real propagation and decay vectors $\mbf{k}_{+}$ and $\mbf{k}_{-}$, we construct the complex propagation vector $\wt{\mbf{k}}$ according to its magnitude $\wt{k}$ and direction $\uvec{k}$ and then decompose its real and imaginary parts, $\mbf{k}_{+} + i \mbf{k}_{-} = \wt{k} \uvec{k}$.  The directions for propagation and decay are given by the unit vectors $\mbf{k}_{+} /  \abs{\mbf{k}_{+}}$ and $\mbf{k}_{-} / \abs{\mbf{k}_{-}}$, which for $\omega > \omega_c$ are found not to point in the same direction as a phase in addition to $\delta_k$ appears when $\uvec{k}$ is complex.  When $\omega \leq \omega_c$ the unit vector $\uvec{k}$ is real, and for all $\omega$ one finds $\uvec{k} \cdot (\mbf{f}_{+} \times \mbf{f}_{-}) = 1$.

\begin{figure}
\includegraphics[scale=0.8]{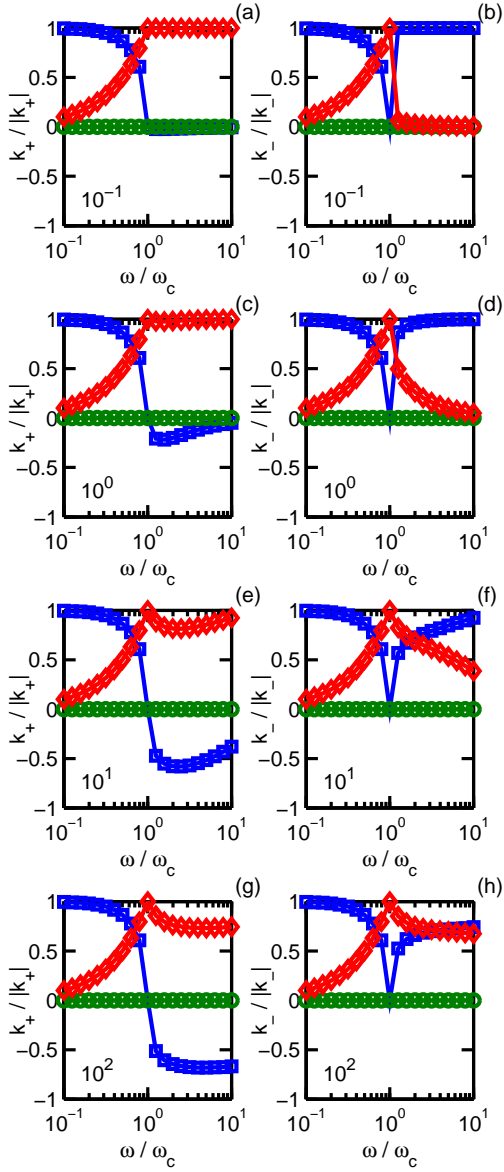}%
\caption{\label{fig:E}(Color online.)  Shown are the unit vectors giving the directions of propagation $\mbf{k}_{+}$ and decay $\mbf{k}_{-}$ as a function of the frequency ratio for various values of $\rho_0$ indicated to the lower left of each panel.  The $(x,y,z)$ components are displayed as $\square$, $\bigcirc$, and $\lozenge$ respectively.}
\end{figure}

For various $\rho_0 \in [10^{-1}, 10^2]$, the directions of propagation and decay as a function of the frequency ratio are shown in Fig.~\ref{fig:E}.  When $\omega \leq \omega_c$ they are both equal to $\uvec{k} = \mbf{f}_{+} \times \mbf{f}_{-}$, which does not depend on $\rho_0$.  For $\omega \gg \omega_c$ they are nearly but not quite orthogonal, as $\mbf{k}_{+} \cdot \mbf{k}_{-} > 0$, and the extent of the transition region does depend on $\rho_0$.  Attempting to describe the behavior, as $\omega$ goes from 0 to $\omega_c$ the propagation and decay vectors both point along the normal to the flow plane, which in this case $\theta_k = 0$ goes from $\uvec{x}$ to $\uvec{z}$ in the $x \mbox{-} z$ plane.  As $\omega$ increases beyond the cyclotron frequency, the propagation direction first acquires a component along $-\uvec{x}$ before swinging back to $\uvec{z}$.  The direction of decay proceeds back towards $\uvec{x}$, and the normal to the flow plane heads towards $\uvec{y}$ in the $y \mbox{-} z$ plane.  As the ratio $\rho_0$ increases, the transition region for the propagation direction grows both in terms of the swing towards $-\uvec{x}$ and how high a frequency ratio is needed before pointing along $\uvec{z}$, and similarly for the direction of decay.

\begin{figure}
\includegraphics[scale=0.8]{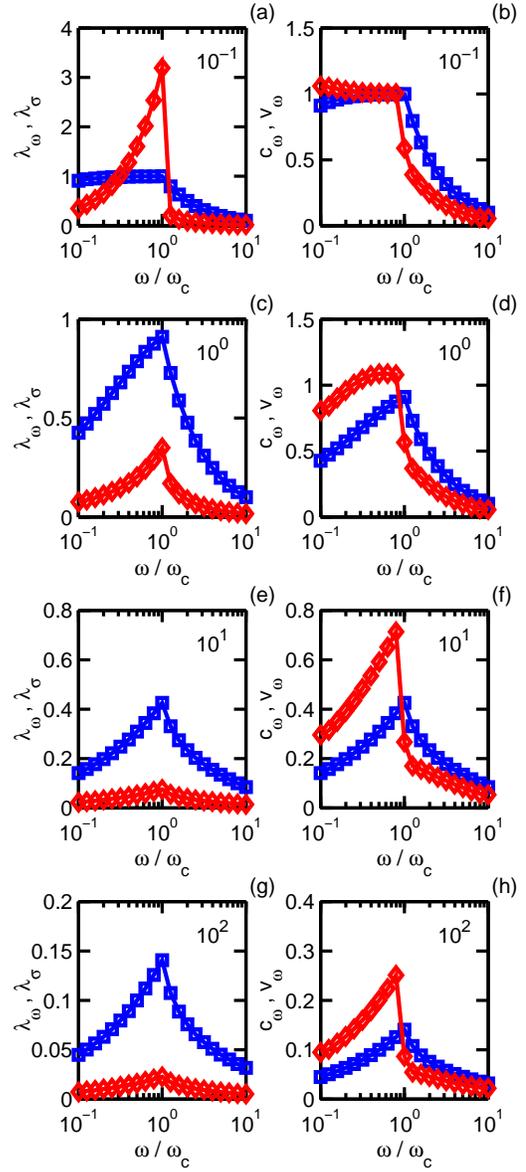}%
\caption{\label{fig:F}(Color online.)  Shown on the left are the wavelength $\lambda_\omega$ as $\square$ and the skin depth $\lambda_\sigma$ as $\lozenge$ in units of the vacuum wavelength $\lambda_0$ as a function of the frequency ratio for various $\rho_0$ indicated to the upper right of each panel.  On the right are the phase speed $c_\omega$ as $\square$ and the group speed $v_\omega$ as $\lozenge$ in units of the vacuum light speed $c_0$.}
\end{figure}

The magnitudes of the propagation and decay vectors determine the wavelength and skin depth as a function of the frequency ratio, $\lambda_\omega = 2 \pi / \abs{\mbf{k}_{+}}$ and $\lambda_\sigma = 1 / \abs{\mbf{k}_{-}}$.  The dispersion relation yields the phase speed $c_\omega = \omega / \abs{\mbf{k}_{+}}$, and its derivative the group speed $v_\omega = \partial \omega / \partial \abs{\mbf{k}_{+}}$.  For all $\omega$ one can decompose the complex propagation vector $\wt{\mbf{k}}$ as \bes
\wt{k} \uvec{k} &\equiv& ( k_{+} + i k_{-} ) ( \uvec{k}_{+} + i \uvec{k}_{-} ) \\
 &=& ( k_{+} \uvec{k}_{+} - k_{-} \uvec{k}_{-} ) + i ( k_{-} \uvec{k}_{+} + k_{+} \uvec{k}_{-} ) \; ,
\ees and when $\uvec{k}_{-} = 0$ one has $\abs{\mbf{k}_{+}} = k_{+}$ and $\abs{\mbf{k}_{-}} = k_{-}$.  In Fig.~\ref{fig:F} we display $\lambda_\omega$ and $\lambda_\sigma$ in units of the vacuum wavelength $\lambda_0 = c_0 / (\omega / 2 \pi)$ as well as $c_\omega$ and $v_\omega$ in units of the vacuum light speed $c_0$, for $\rho_0 \in [10^{-1}, 10^2]$.

The transition between conductive and non-conductive behavior occurs for a value of $\rho_0 \lesssim 1$, such that $\lambda_\sigma > \lambda_\omega$ for a range of frequencies $\omega < \omega_c$.  In other words, significant propagation of the oscillation over several wavelengths occurs only for a span of frequencies below the cyclotron resonance when the conductivity is sufficiently low that dissipation does not destroy the waveform.  For higher conductivities the skin depth is but a fraction of the wavelength, indicating that the amplitude decays to almost nothing before a single cycle is realized.  For $\omega < \omega_c$ the phase speed is less than the group speed, and for $\omega > \omega_c$ it is greater.  Interestingly, for a poor conductor $\rho_0 \lesssim 1$ there exists a range of frequencies below $\omega_c$ for which the group speed exceeds the vacuum light speed.  Such a case is not unheard of in plasma physics,\cite{peters-129} and we stress that this result obtains directly from the accepted solution to the inhomogeneous d'Alembertian equation for conductors.\cite{griffiths-89}  One can verify that the group speeds displayed, evaluated from numerical gradients, agree with the analytic expression \beq
v_{\omega < \omega_c} = \dfrac{2 ( \rho^2 + 1 )^{3/4} c_0}{\rho \sin \delta_k + ( \rho^2 + 2 ) \cos \delta_k} \; , \label{eqn:vglo} 
\eeq for frequencies below the cyclotron resonance, and above the resonance \bew \beq
v_{\omega > \omega_c} = \dfrac {2 ( \rho^2 + 1 )^{3/4} \left [ ( \omega / \omega_c )^2 - \sin^2 \delta_k \right ]^{1/2} c_0} {\rho \cos \delta_k \sin \delta_k - ( \rho^2 + 2 ) \sin^2 \delta_k + ( \omega / \omega_c )^2 ( 3 \rho^2 + 4 )} \; , \label{eqn:vghi}
\eeq \eew recalling $\rho = \tan (2 \delta_k)$.  Equation~(\ref{eqn:vglo}) can be derived from textbook electrodynamics, while Eqn.~(\ref{eqn:vghi}) is a contribution of this investigation taking into account $\uvec{k}_{-}$.

\section{Discussion}
In discussing the significance of these results, let us begin by enumerating some of the many effects neglected by this analysis.  By using the vacuum values for permittivity and permeability $\epsi_0$ and $\mu_0$, one has ignored both the electric and magnetic atomic polarizability of the ions.  Obviously these effects will have an impact on the evaluation of the wave number $\wt{k}$ and all quantities derived therefrom.  For that reason we are not overly concerned by the appearance of a group speed in excess of the vacuum light speed at this stage in the development of the theory.  When the atomic polarizabilities can by absorbed into scalar relative permittivity and permeability $\epsi$ and $\mu$, one simply replaces the vacuum expressions with those values; however, in a magnetized plasma, those quantities are usually defined as gyrotropic tensors derived from the equations of motion.

The motion of the ions is neglected throughout these derivations.  To the extent that the ion and electron motions decouple, one can repeat the analysis for $\mbf{J} = e n_i \mbf{V}_i$ over a range of frequencies around the ion cyclotron frequency to yield a ``slow'' partner to the ``fast'' mode derived above for the electrons.  As the ions (are assumed to) carry positive charge, their motion couples to electromagnetic oscillations of opposite circular polarization to that for electrons.\cite{wallace-196}  The most interesting case of course is when both electrons and ions are allowed to flow.  In that situation their combined flow can be decomposed into net current and momentum oscillations $\wt{\mbf{J}}$ and $\wt{\mbf{K}}$ coupled through the equations of motion.  One expects that low frequencies will excite mostly momentum waves, whereas high frequencies should drive mostly current waves, with separate transition frequencies for positive and negative helical polarization.

An important effect neglected here is the possibility of temporal and spatial variation to the electron density, as well as the presence of sources or sinks of current, $(\dsub{t} + \mbf{V}_e \dotdel) n_e = \dot{n}_e - (\divr \mbf{V}_e) n_e$.  It is the effect of a net charge density, after all, which pushes current around a source driven wire; Hernandes and Assis\cite{PRE-046611} give an example of its derivation in the steady state and Jefimenko\cite{jefimenko-19} gives an example of its measurement.  In Lorenz gauge, the net charge density $j$ couples to the scalar potential $a$ through the same d'Alembertian operator $- \dalem a = \mu_0 j$.  With respect to the derivation of the (electron) plasma frequency $\omega_p = n_e e^2 m_e \epsi_0$, a non-vanishing $j$ should allow for the coupling of the electron fluid to electromagnetic oscillations of linear polarization in addition to the circular polarization described here.  In this context, it might be beneficial to decompose the statement of net momentum conservation Eqn.~(\ref{eqn:netforce}) in terms of the Maxwell stress tensor\cite{mansuripur-1619,rwj-jpp03} so that the components of the macroscopic Lorentz force can be identified.

As the only external potential apparent in the theory is that of the background field $\mbf{B}_0$, what we have derived is essentially a model for resonance within the electron fluid, not its response to a driving potential at arbitrary frequency.  In order to describe the plasma response to an external device, such as a radio frequency antenna, the potential from that device must be calculated and included explicitly in the mathematics.  For the common configuration of a plasma confined by a cylindrical vessel,\cite{chen-123507,palmer-0609,kwon-045021} one also must be explicit with the transition from conductive to dielectric material at the boundary such that $n_e \rightarrow 0$.  Nonetheless, the model derived here is based upon the same simplifying approximations as commonly used in the theoretical description of such devices\cite{chen-173,chen-174,yano-063501,yano-033510} and sheds some light on the electromagnetic behavior of ionized material.

What we have learned is that one need not neglect the nonlinear term in the equation of motion to find an analytic solution when the ion contribution is included as a constraint relating the electron flow to the potential it generates.  Working with the electromagnetic potential, neither is Gauss's law neglected; rather, it is the possible occurrence of a net charge density which is neglected in the neutral fluid approximation.  Our main result is that propagation occurs in an ionized medium under the given conditions only for a range of frequencies below the electron cyclotron resonance when the conductivity is sufficiently low that the displacement current outweighs the conduction current.  For a highly conductive medium, the skin depth dominates the wavelength such that the amplitude of oscillation is attenuated.  Below the cyclotron resonance the propagation and decay vectors point along the normal to the flow plane, whereas above the resonance they point in different directions while still respecting continuity.

\section{Conclusion}
In closing, we hope that this article has contributed to the understanding of the analysis of dispersion in an ionized medium subject to a background magnetic field.  The solution presented here satisfies the classical Maxwell equations, which may be expressed succinctly in terms of geometric forms\cite{ryder-qft,naka-798212} as ${\mrm{d} \star \mrm{d} A = J}$ for ${\mrm{d} \mrm{d} A \equiv 0}$, as well as both the ion and electron equations of motion under the given simplifying approximations.  From the plane wave expansion one should be able to construct solutions of arbitrary geometry.  The nonlinear coupling of the electron flow to its own potential leads to a model which specifies a relation between the frequency of oscillation and the inclination of the propagation vector from the plane normal to the background field.  The circulatory flow given by the vector phasor describes the magnetization of the electron fluid in the presence of a propagating electromagnetic oscillation.

To describe systems of physical interest, this model must be extended to a multispecies formalism which allows the ions to flow, incorporates the effects of intraspecies collisions through the thermal stress tensors, and accounts for a net charge density.  Nonetheless, this simplified model reveals some interesting physics, namely that the propagation of electromagnetic radiation in a uniform plasma subject to a constant magnetic field induces a rotation of the magnetization away from the axis of the background field.  The ratio of oscillation and cyclotron frequencies determines the inclination of the propagation vector, which is orthogonal to the plane of flow when that ratio is less than unity.  The solution decays in space according to the ratio of free to displacement current, which depends upon the conductivity of the medium and the frequency of oscillation.

The motivation for this investigation is the question of whether helicon waves really exist, to which our answer is yes, they do exist, just not the ones given by previous derivations.  Working in the potential formulation simplifies the analysis by reducing the electromagnetic sector to its physical degrees of freedom, so that the nonlinear self-interaction between the plasma current and the potential it generates can be addressed in the equation of motion.  Under the stated conditions, an analytic solution can be found to the nonlinear system of equations describing resonance within the electron fluid of a uniform, stationary plasma subject to a constant background magnetic field.

%\begin{acknowledgments}
% Put your acknowledgments here.
%\end{acknowledgments}

%%%%  END LINE NUMBERS
\makeatletter
\@ifundefined{linenumbers}{}{
\nolinenumbers
}
\makeatother

%\newpage

%\bibliography{../plasma}

\begin{thebibliography}{29}%
\makeatletter
\providecommand \@ifxundefined [1]{%
 \@ifx{#1\undefined}
}%
\providecommand \@ifnum [1]{%
 \ifnum #1\expandafter \@firstoftwo
 \else \expandafter \@secondoftwo
 \fi
}%
\providecommand \@ifx [1]{%
 \ifx #1\expandafter \@firstoftwo
 \else \expandafter \@secondoftwo
 \fi
}%
\providecommand \natexlab [1]{#1}%
\providecommand \enquote  [1]{``#1''}%
\providecommand \bibnamefont  [1]{#1}%
\providecommand \bibfnamefont [1]{#1}%
\providecommand \citenamefont [1]{#1}%
\providecommand \href@noop [0]{\@secondoftwo}%
\providecommand \href [0]{\begingroup \@sanitize@url \@href}%
\providecommand \@href[1]{\@@startlink{#1}\@@href}%
\providecommand \@@href[1]{\endgroup#1\@@endlink}%
\providecommand \@sanitize@url [0]{\catcode `\\12\catcode `\$12\catcode
  `\&12\catcode `\#12\catcode `\^12\catcode `\_12\catcode `\%12\relax}%
\providecommand \@@startlink[1]{}%
\providecommand \@@endlink[0]{}%
\providecommand \url  [0]{\begingroup\@sanitize@url \@url }%
\providecommand \@url [1]{\endgroup\@href {#1}{\urlprefix }}%
\providecommand \urlprefix  [0]{URL }%
\providecommand \Eprint [0]{\href }%
\providecommand \doibase [0]{http://dx.doi.org/}%
\providecommand \selectlanguage [0]{\@gobble}%
\providecommand \bibinfo  [0]{\@secondoftwo}%
\providecommand \bibfield  [0]{\@secondoftwo}%
\providecommand \translation [1]{[#1]}%
\providecommand \BibitemOpen [0]{}%
\providecommand \bibitemStop [0]{}%
\providecommand \bibitemNoStop [0]{.\EOS\space}%
\providecommand \EOS [0]{\spacefactor3000\relax}%
\providecommand \BibitemShut  [1]{\csname bibitem#1\endcsname}%
\let\auto@bib@innerbib\@empty
%</preamble>
\bibitem [{\citenamefont {{Chen}}(1991)}]{chen-137}%
  \BibitemOpen
  \bibfield  {author} {\bibinfo {author} {\bibfnamefont {F.~F.}\ \bibnamefont
  {{Chen}}},\ }\href@noop {} {\bibfield  {journal} {\bibinfo  {journal} {Plasma
  Phys. Control. Fusion}\ }\textbf {\bibinfo {volume} {33}},\ \bibinfo {pages}
  {339} (\bibinfo {year} {1991})}\BibitemShut {NoStop}%
\bibitem [{\citenamefont {Trivelpiece}\ and\ \citenamefont
  {Gould}(1959)}]{trivelpiece-1784}%
  \BibitemOpen
  \bibfield  {author} {\bibinfo {author} {\bibfnamefont {A.~W.}\ \bibnamefont
  {Trivelpiece}}\ and\ \bibinfo {author} {\bibfnamefont {R.~W.}\ \bibnamefont
  {Gould}},\ }\href {\doibase 10.1063/1.1735056} {\bibfield  {journal}
  {\bibinfo  {journal} {Journal of Applied Physics}\ }\textbf {\bibinfo
  {volume} {30}},\ \bibinfo {pages} {1784} (\bibinfo {year}
  {1959})}\BibitemShut {NoStop}%
\bibitem [{\citenamefont {Bowers}, \citenamefont {Legendy},\ and\ \citenamefont
  {Rose}(1961)}]{bowers-1961}%
  \BibitemOpen
  \bibfield  {author} {\bibinfo {author} {\bibfnamefont {R.}~\bibnamefont
  {Bowers}}, \bibinfo {author} {\bibfnamefont {C.}~\bibnamefont {Legendy}}, \
  and\ \bibinfo {author} {\bibfnamefont {F.}~\bibnamefont {Rose}},\ }\href
  {\doibase 10.1103/PhysRevLett.7.339} {\bibfield  {journal} {\bibinfo
  {journal} {Phys. Rev. Lett.}\ }\textbf {\bibinfo {volume} {7}},\ \bibinfo
  {pages} {339} (\bibinfo {year} {1961})}\BibitemShut {NoStop}%
\bibitem [{\citenamefont {Leg\'endy}(1964)}]{legendy-1964}%
  \BibitemOpen
  \bibfield  {author} {\bibinfo {author} {\bibfnamefont {C.~R.}\ \bibnamefont
  {Leg\'endy}},\ }\href {\doibase 10.1103/PhysRev.135.A1713} {\bibfield
  {journal} {\bibinfo  {journal} {Phys. Rev.}\ }\textbf {\bibinfo {volume}
  {135}},\ \bibinfo {pages} {A1713} (\bibinfo {year} {1964})}\BibitemShut
  {NoStop}%
\bibitem [{\citenamefont {Klozenberg}, \citenamefont {McNamara},\ and\
  \citenamefont {Thonemann}(1965)}]{KMT-1965}%
  \BibitemOpen
  \bibfield  {author} {\bibinfo {author} {\bibfnamefont {J.~P.}\ \bibnamefont
  {Klozenberg}}, \bibinfo {author} {\bibfnamefont {B.}~\bibnamefont
  {McNamara}}, \ and\ \bibinfo {author} {\bibfnamefont {P.~C.}\ \bibnamefont
  {Thonemann}},\ }\href {\doibase 10.1017/S0022112065000320} {\bibfield
  {journal} {\bibinfo  {journal} {Journal of Fluid Mechanics}\ }\textbf
  {\bibinfo {volume} {21}},\ \bibinfo {pages} {545} (\bibinfo {year}
  {1965})}\BibitemShut {NoStop}%
\bibitem [{\citenamefont {{Chen}}(1995)}]{chen-155r}%
  \BibitemOpen
  \bibfield  {author} {\bibinfo {author} {\bibfnamefont {F.~F.}\ \bibnamefont
  {{Chen}}},\ }in\ \href@noop {} {\emph {\bibinfo {booktitle} {High Density
  Plasma Sources}}},\ \bibinfo {editor} {edited by\ \bibinfo {editor}
  {\bibfnamefont {O.~A.}\ \bibnamefont {Popov}}}\ (\bibinfo  {publisher} {Noyes
  Publications},\ \bibinfo {address} {Park Ridge, NJ},\ \bibinfo {year}
  {1995})\BibitemShut {NoStop}%
\bibitem [{\citenamefont {Boswell}\ and\ \citenamefont
  {Chen}(1997)}]{chen-173}%
  \BibitemOpen
  \bibfield  {author} {\bibinfo {author} {\bibfnamefont {R.~W.}\ \bibnamefont
  {Boswell}}\ and\ \bibinfo {author} {\bibfnamefont {F.~F.}\ \bibnamefont
  {Chen}},\ }\href {\doibase 10.1109/27.650898} {\bibfield  {journal} {\bibinfo
   {journal} {IEEE Trans. Plasma Sci.}\ }\textbf {\bibinfo {volume} {25}},\
  \bibinfo {pages} {1229} (\bibinfo {year} {1997})}\BibitemShut {NoStop}%
\bibitem [{\citenamefont {Chen}\ and\ \citenamefont
  {Boswell}(1997)}]{chen-174}%
  \BibitemOpen
  \bibfield  {author} {\bibinfo {author} {\bibfnamefont {F.~F.}\ \bibnamefont
  {Chen}}\ and\ \bibinfo {author} {\bibfnamefont {R.~W.}\ \bibnamefont
  {Boswell}},\ }\href {\doibase 10.1109/27.650899} {\bibfield  {journal}
  {\bibinfo  {journal} {IEEE Trans. Plasma Sci.}\ }\textbf {\bibinfo {volume}
  {25}},\ \bibinfo {pages} {1245} (\bibinfo {year} {1997})}\BibitemShut
  {NoStop}%
\bibitem [{\citenamefont {Tripathi}\ and\ \citenamefont
  {Bora}(2001)}]{tripathi-697}%
  \BibitemOpen
  \bibfield  {author} {\bibinfo {author} {\bibfnamefont {S.~K.~P.}\
  \bibnamefont {Tripathi}}\ and\ \bibinfo {author} {\bibfnamefont
  {D.}~\bibnamefont {Bora}},\ }\href {\doibase 10.1063/1.1344198} {\bibfield
  {journal} {\bibinfo  {journal} {Physics of Plasmas}\ }\textbf {\bibinfo
  {volume} {8}},\ \bibinfo {pages} {697} (\bibinfo {year} {2001})}\BibitemShut
  {NoStop}%
\bibitem [{\citenamefont {Yano}\ and\ \citenamefont
  {Walker}(2006)}]{yano-063501}%
  \BibitemOpen
  \bibfield  {author} {\bibinfo {author} {\bibfnamefont {M.}~\bibnamefont
  {Yano}}\ and\ \bibinfo {author} {\bibfnamefont {M.~L.~R.}\ \bibnamefont
  {Walker}},\ }\href {\doibase 10.1063/1.2207125} {\bibfield  {journal}
  {\bibinfo  {journal} {Physics of Plasmas}\ }\textbf {\bibinfo {volume}
  {13}},\ \bibinfo {eid} {063501} (\bibinfo {year} {2006})}\BibitemShut
  {NoStop}%
\bibitem [{\citenamefont {Yano}\ and\ \citenamefont
  {Walker}(2007)}]{yano-033510}%
  \BibitemOpen
  \bibfield  {author} {\bibinfo {author} {\bibfnamefont {M.}~\bibnamefont
  {Yano}}\ and\ \bibinfo {author} {\bibfnamefont {M.~L.~R.}\ \bibnamefont
  {Walker}},\ }\href {\doibase 10.1063/1.2716663} {\bibfield  {journal}
  {\bibinfo  {journal} {Physics of Plasmas}\ }\textbf {\bibinfo {volume}
  {14}},\ \bibinfo {eid} {033510} (\bibinfo {year} {2007})}\BibitemShut
  {NoStop}%
\bibitem [{\citenamefont {{Jankauskas}}\ and\ \citenamefont
  {{Kvedaras}}(2007)}]{jankauskas-274}%
  \BibitemOpen
  \bibfield  {author} {\bibinfo {author} {\bibfnamefont {Z.}~\bibnamefont
  {{Jankauskas}}}\ and\ \bibinfo {author} {\bibfnamefont {V.}~\bibnamefont
  {{Kvedaras}}},\ }\href@noop {} {\bibfield  {journal} {\bibinfo  {journal}
  {Electronics and Electrical Engineering}\ }\textbf {\bibinfo {volume} {2}},\
  \bibinfo {pages} {41} (\bibinfo {year} {2007})}\BibitemShut {NoStop}%
\bibitem [{\citenamefont {Chen}\ and\ \citenamefont {Arnush}(1997)}]{chen-175}%
  \BibitemOpen
  \bibfield  {author} {\bibinfo {author} {\bibfnamefont {F.~F.}\ \bibnamefont
  {Chen}}\ and\ \bibinfo {author} {\bibfnamefont {D.}~\bibnamefont {Arnush}},\
  }\href {\doibase 10.1063/1.872483} {\bibfield  {journal} {\bibinfo  {journal}
  {Physics of Plasmas}\ }\textbf {\bibinfo {volume} {4}},\ \bibinfo {pages}
  {3411} (\bibinfo {year} {1997})}\BibitemShut {NoStop}%
\bibitem [{\citenamefont {{Griffiths}}(1989)}]{griffiths-89}%
  \BibitemOpen
  \bibfield  {author} {\bibinfo {author} {\bibfnamefont {D.}~\bibnamefont
  {{Griffiths}}},\ }\href@noop {} {\emph {\bibinfo {title} {Introduction to
  Electrodynamics}}},\ \bibinfo {edition} {2nd}\ ed.\ (\bibinfo  {publisher}
  {{Prentice-Hall, Inc.}},\ \bibinfo {address} {Englewood Cliffs, NJ},\
  \bibinfo {year} {1989})\BibitemShut {NoStop}%
\bibitem [{\citenamefont {{Dendy}}(1993)}]{dendybook-93}%
  \BibitemOpen
  \bibinfo {editor} {\bibfnamefont {R.}~\bibnamefont {{Dendy}}},\ ed.,\
  \href@noop {} {\emph {\bibinfo {title} {Plasma Physics: an Introductory
  Course}}}\ (\bibinfo  {publisher} {Cambridge University Press},\ \bibinfo
  {address} {Cambridge, UK},\ \bibinfo {year} {1993})\BibitemShut {NoStop}%
\bibitem [{\citenamefont {{Ryder}}(1985)}]{ryder-qft}%
  \BibitemOpen
  \bibfield  {author} {\bibinfo {author} {\bibfnamefont {L.~H.}\ \bibnamefont
  {{Ryder}}},\ }\href@noop {} {\emph {\bibinfo {title} {Quantum Field
  Theory}}}\ (\bibinfo  {publisher} {Cambridge University Press},\ \bibinfo
  {address} {Cambridge, UK},\ \bibinfo {year} {1985})\BibitemShut {NoStop}%
\bibitem [{\citenamefont {Nakahara}(1990)}]{naka-798212}%
  \BibitemOpen
  \bibfield  {author} {\bibinfo {author} {\bibfnamefont {M.}~\bibnamefont
  {Nakahara}},\ }\href@noop {} {\emph {\bibinfo {title} {Geometry, Topology and
  Physics}}}\ (\bibinfo  {publisher} {IOP Publishing Ltd.},\ \bibinfo {address}
  {Bristol, UK},\ \bibinfo {year} {1990})\BibitemShut {NoStop}%
\bibitem [{\citenamefont {Bork}(1963)}]{bork-854}%
  \BibitemOpen
  \bibfield  {author} {\bibinfo {author} {\bibfnamefont {A.~M.}\ \bibnamefont
  {Bork}},\ }\href {\doibase 10.1119/1.1969140} {\bibfield  {journal} {\bibinfo
   {journal} {American Journal of Physics}\ }\textbf {\bibinfo {volume} {31}},\
  \bibinfo {pages} {854} (\bibinfo {year} {1963})}\BibitemShut {NoStop}%
\bibitem [{\citenamefont {Stacey}(2005)}]{staceybook05}%
  \BibitemOpen
  \bibfield  {author} {\bibinfo {author} {\bibfnamefont {W.~M.}\ \bibnamefont
  {Stacey}},\ }\href@noop {} {\emph {\bibinfo {title} {Fusion Plasma
  Physics}}}\ (\bibinfo  {publisher} {Wiley-VCH},\ \bibinfo {address} {New
  York, NY},\ \bibinfo {year} {2005})\BibitemShut {NoStop}%
\bibitem [{\citenamefont {Flanigan}(1983)}]{flanigan-1983}%
  \BibitemOpen
  \bibfield  {author} {\bibinfo {author} {\bibfnamefont {F.}~\bibnamefont
  {Flanigan}},\ }\href {http://books.google.com/books?id=X0ILoaB7LREC} {\emph
  {\bibinfo {title} {Complex Variables: Harmonic and Analytic Functions}}},\
  Dover books on advanced mathematics\ (\bibinfo  {publisher} {Dover
  Publications},\ \bibinfo {year} {1983})\BibitemShut {NoStop}%
\bibitem [{\citenamefont {Peters}(1988)}]{peters-129}%
  \BibitemOpen
  \bibfield  {author} {\bibinfo {author} {\bibfnamefont {P.~C.}\ \bibnamefont
  {Peters}},\ }\href {\doibase 10.1119/1.15715} {\bibfield  {journal} {\bibinfo
   {journal} {American Journal of Physics}\ }\textbf {\bibinfo {volume} {56}},\
  \bibinfo {pages} {129} (\bibinfo {year} {1988})}\BibitemShut {NoStop}%
\bibitem [{\citenamefont {Wallace}(1964)}]{wallace-196}%
  \BibitemOpen
  \bibfield  {author} {\bibinfo {author} {\bibfnamefont {P.~R.}\ \bibnamefont
  {Wallace}},\ }\href {\doibase 10.1139/p64-196} {\bibfield  {journal}
  {\bibinfo  {journal} {Canadian Journal of Physics}\ }\textbf {\bibinfo
  {volume} {42}},\ \bibinfo {pages} {2129} (\bibinfo {year}
  {1964})}\BibitemShut {NoStop}%
\bibitem [{\citenamefont {Hernandes}\ and\ \citenamefont
  {Assis}(2003)}]{PRE-046611}%
  \BibitemOpen
  \bibfield  {author} {\bibinfo {author} {\bibfnamefont {J.~A.}\ \bibnamefont
  {Hernandes}}\ and\ \bibinfo {author} {\bibfnamefont {A.~K.~T.}\ \bibnamefont
  {Assis}},\ }\href {\doibase 10.1103/PhysRevE.68.046611} {\bibfield  {journal}
  {\bibinfo  {journal} {Phys. Rev. E}\ }\textbf {\bibinfo {volume} {68}},\
  \bibinfo {pages} {046611} (\bibinfo {year} {2003})}\BibitemShut {NoStop}%
\bibitem [{\citenamefont {Jefimenko}(1962)}]{jefimenko-19}%
  \BibitemOpen
  \bibfield  {author} {\bibinfo {author} {\bibfnamefont {O.}~\bibnamefont
  {Jefimenko}},\ }\href {\doibase 10.1119/1.1941887} {\bibfield  {journal}
  {\bibinfo  {journal} {American Journal of Physics}\ }\textbf {\bibinfo
  {volume} {30}},\ \bibinfo {pages} {19} (\bibinfo {year} {1962})}\BibitemShut
  {NoStop}%
\bibitem [{\citenamefont {Mansuripur}(2008)}]{mansuripur-1619}%
  \BibitemOpen
  \bibfield  {author} {\bibinfo {author} {\bibfnamefont {M.}~\bibnamefont
  {Mansuripur}},\ }\href
  {http://www.opticsexpress.org/abstract.cfm?URI=oe-16-19-14821} {\bibfield
  {journal} {\bibinfo  {journal} {Opt. Express}\ }\textbf {\bibinfo {volume}
  {16}},\ \bibinfo {pages} {14821} (\bibinfo {year} {2008})}\BibitemShut
  {NoStop}%
\bibitem [{\citenamefont {{Johnson}}(2011)}]{rwj-jpp03}%
  \BibitemOpen
  \bibfield  {author} {\bibinfo {author} {\bibfnamefont {R.~W.}\ \bibnamefont
  {{Johnson}}},\ }\href {\doibase 10.1017/S002237780999050X} {\bibfield
  {journal} {\bibinfo  {journal} {Journal of Plasma Physics}\ }\textbf
  {\bibinfo {volume} {77}},\ \bibinfo {pages} {107} (\bibinfo {year}
  {2011})}\BibitemShut {NoStop}%
\bibitem [{\citenamefont {Chen}\ \emph {et~al.}(2006)\citenamefont {Chen},
  \citenamefont {Arefiev}, \citenamefont {Bengtson}, \citenamefont {Breizman},
  \citenamefont {Lee},\ and\ \citenamefont {Raja}}]{chen-123507}%
  \BibitemOpen
  \bibfield  {author} {\bibinfo {author} {\bibfnamefont {G.}~\bibnamefont
  {Chen}}, \bibinfo {author} {\bibfnamefont {A.~V.}\ \bibnamefont {Arefiev}},
  \bibinfo {author} {\bibfnamefont {R.~D.}\ \bibnamefont {Bengtson}}, \bibinfo
  {author} {\bibfnamefont {B.~N.}\ \bibnamefont {Breizman}}, \bibinfo {author}
  {\bibfnamefont {C.~A.}\ \bibnamefont {Lee}}, \ and\ \bibinfo {author}
  {\bibfnamefont {L.~L.}\ \bibnamefont {Raja}},\ }\href {\doibase
  10.1063/1.2402913} {\bibfield  {journal} {\bibinfo  {journal} {Physics of
  Plasmas}\ }\textbf {\bibinfo {volume} {13}},\ \bibinfo {eid} {123507}
  (\bibinfo {year} {2006})}\BibitemShut {NoStop}%
\bibitem [{\citenamefont {{Palmer}}\ and\ \citenamefont
  {{Walker}}(2009)}]{palmer-0609}%
  \BibitemOpen
  \bibfield  {author} {\bibinfo {author} {\bibfnamefont {D.~D.}\ \bibnamefont
  {{Palmer}}}\ and\ \bibinfo {author} {\bibfnamefont {M.~L.~R.}\ \bibnamefont
  {{Walker}}},\ }\href@noop {} {\bibfield  {journal} {\bibinfo  {journal}
  {Journal of Propulsion and Power}\ }\textbf {\bibinfo {volume} {25}},\
  \bibinfo {pages} {1013} (\bibinfo {year} {2009})}\BibitemShut {NoStop}%
\bibitem [{\citenamefont {Kwon}, \citenamefont {Walker},\ and\ \citenamefont
  {Mavris}(2011)}]{kwon-045021}%
  \BibitemOpen
  \bibfield  {author} {\bibinfo {author} {\bibfnamefont {K.}~\bibnamefont
  {Kwon}}, \bibinfo {author} {\bibfnamefont {M.~L.}\ \bibnamefont {Walker}}, \
  and\ \bibinfo {author} {\bibfnamefont {D.~N.}\ \bibnamefont {Mavris}},\
  }\href {http://stacks.iop.org/0963-0252/20/i=4/a=045021} {\bibfield
  {journal} {\bibinfo  {journal} {Plasma Sources Science and Technology}\
  }\textbf {\bibinfo {volume} {20}},\ \bibinfo {pages} {045021} (\bibinfo
  {year} {2011})}\BibitemShut {NoStop}%
\end{thebibliography}


%%% RESPONSE
%\newpage
%\input{response04}

\end{document}